\documentclass[10pt,twocolumn,preprintnumbers,superscriptaddress,nofootinbib,aps,prl]{revtex4}

\usepackage{graphicx}
\usepackage{amsmath}
\usepackage{srcltx}
\usepackage[utf8]{inputenc}
\usepackage[colorlinks]{hyperref}
\usepackage{times}
\usepackage{subfigure}
\newcommand{\vev}[1]{\langle {#1} \rangle}
\newcommand{\lsim}{\lesssim}
\newcommand{\gsim}{\gtrsim}
\newcommand{\avsig}{\vev{\sigma_{\rm wo}}}
\newcommand{\eq}[1]{Eq.~(\ref{#1})}

\newcommand{\ord}[1]{\mathcal{O}{(#1)}}
\newcommand{\beq}{\begin{equation}}
\newcommand{\eeq}{\end{equation}}
\newcommand{\eps}{\varepsilon}

\newcommand{\mP}{M_{\rm P}}

\newcommand{\DBL}{\Delta (B-L)}

\newcommand{\appropto}{\mathrel{\vcenter{
  \offinterlineskip\halign{\hfil$##$\cr
    \propto\cr\noalign{\kern2pt}\sim\cr\noalign{\kern-2pt}}}}}

\begin{document}

\pagestyle{plain}

\title{\boldmath Multi-TeV Signals of Baryogenesis in Higgs Troika Model}

\author{Hooman Davoudiasl\footnote{email: hooman@bnl.gov}
}

\affiliation{High Energy Theory Group, Physics Department, Brookhaven National Laboratory,
Upton, New York 11973, USA}

\author{Ian M. Lewis\footnote{email: ian.lewis@ku.edu}
}

\affiliation{Department of Physics and Astronomy, University of Kansas, 
Lawrence, Kansas, 66045 USA}

\author{Matthew Sullivan\footnote{email: msullivan1@bnl.gov
}
}

\affiliation{High Energy Theory Group, Physics Department, Brookhaven National Laboratory,
Upton, New York 11973, USA}


\begin{abstract}

A modest extension of the Standard Model by two additional Higgs doublets - the {\it Higgs Troika Model} - can provide a well-motivated scenario for successful baryogenesis if neutrinos are Dirac fermions.  Adapting the ``Spontaneous Flavor Violation'' framework, we consider a version of the Troika model where light quarks have significant couplings to the new multi-TeV Higgs states.  Resonant production of new scalars leading to di-jet or top-pair signals are typical predictions of this setup.  The initial and final state quarks relevant to the collider phenomenology also play a key role in baryogenesis, potentially providing direct access to the relevant early Universe physics in high energy experiments.  Viable baryogenesis generally prefers some hierarchy of masses between the observed and the postulated Higgs states.  We show that there is a complementarity between direct searches at a future 100 TeV$pp$ collider and indirect searches at flavor experiments, with both sensitive to different regions of parameter space relevant for baryogenesis. In particular, measurements of $D-\bar{D}$ mixing at LHCb probe much of the interesting parameter space.  Direct and indirect searches can uncover the new Higgs states up to masses of $\ord{10}$~TeV, thereby providing an impressive reach to investigate this model.   
\end{abstract}
\maketitle


\section{Introduction}

The reason why the visible content of the Universe exists at all still eludes a definitive answer.  This is equivalent to asking what gave rise to the observed baryon asymmetry of the Universe (BAU) 
\cite{Tanabashi:2018oca} 
\beq
\frac{n_B}{s} \approx 9\times 10^{-11}\,,
\label{obs-BAU}
\eeq
where $n_B$ is the net baryon number density and $s$ is the cosmic entropy density.  The answer to this question 
is generally believed to be furnished by physics beyond the Standard Model (SM).  This expectation is implied by the criteria, {\it i.e.} the Sakharov conditions \cite{Sakharov:1967dj}, necessary for a successful baryogenesis mechanism that provides the BAU: (i) Baryon number $B$ violation, (ii) C and CP violation, and (iii) departure from equilibrium.  The SM can meet condition (i) through thermal  
electroweak processes often referred to as {\it sphalerons} at temperatures $T\gsim 100$~GeV before electroweak symmetry was broken.  However, conditions (ii) and (iii) are not met to the requisite levels in the SM and hence extensions of it that lead to viable baryogenesis are well-motivated, with numerous 
ideas having been put forward over the years.  

An interesting aspect of baryogenesis through SM sphalerons is its connection 
with the physics of leptons.   That is, these processes can generate the BAU by transforming a primordial asymmetry in $B-L$, where $L$ denotes lepton number.  In particular, if a sufficient amount of lepton asymmetry $\Delta L$ is present at $T\gsim 100$~GeV, the sphalerons can provide the 
observed $\Delta B$.  An interesting possibility is offered by leptogenesis \cite{Fukugita:1986hr} through the decay of heavy 
right-handed Majorana neutrinos, leading to $\Delta L\neq 0$ assuming enough CP violation is present in its interactions.  These states are well-motivated components of the seesaw mechanism  \cite{GellMann:1980vs,Minkowski:1977sc,Mohapatra:1979ia,Ramond:1979py} that leads to light Majorana neutrinos.  However, generic Majorana leptogenesis employs heavy right-handed states 
with masses several orders of magnitude above the weak scale, well beyond the reach of direct experimental measurements.  This limits their signature mainly to indirect evidence from observation of lepton number violation in neutrinoless double beta decay, which has so far yielded null results.                      

In this paper, we consider an extension of the SM that could lead to a viable baryogenesis mechanism, while having potentially observable direct signals at high energy collider experiments.  Our mechanism requires the addition of only two extra Higgs doublets, with the same quantum numbers as the SM Higgs doublet, assuming that the SM has already been augmented with right handed neutrinos that are  very likely necessary to endow SM neutrinos with their observed small masses $\lsim 0.1$~eV.  
In fact, as will be discussed below, we find that this {\it Higgs Troika Model}\footnote{{\it Troika} = A group of three.} \cite{Davoudiasl:2019lcg}  is quite well-motivated if the SM neutrinos are Dirac fermions and seesaw-inspired leptogenesis is not relevant.  The Troika model  can be viewed as a minimal implementation of similar ideas in Refs.~\cite{Dick:1999je,Murayama:2002je,Davoudiasl:2011aa}. For more general discussions of three Higgs doublet models please see Refs.~\cite{Cheng:1987rs,Grossman:1994jb,Cree:2011uy,Ivanov:2011ae,Keus:2013hya,Bento:2017eti,Alves:2020brq,Logan:2020mdz} and references therein.

In light of the above, here we will assume that neutrinos are Dirac fermions, unlike in our prior work  \cite{Davoudiasl:2019lcg}, where right-handed neutrinos were taken to be TeV scale Majorana states.  
In the present work, we also take a different approach to flavor from that adopted in Ref.~\cite{Davoudiasl:2019lcg}, where the new Higgs states were assumed responsible for setting up the lepton sector flavor structure.  Here, in contrast, all fermion masses are taken to get sourced by one Higgs doublet which corresponds to the SM-like scalar observed at the LHC with a mass of $\approx 125$~GeV.  

In the following, we will allow the new TeV scale Higgs states to couple with significant strength to light and heavy quarks, thereby providing a possible resonant production channel through quark initial states~\cite{Altmannshofer:2016zrn,Egana-Ugrinovic:2018znw,Egana-Ugrinovic:2019dqu}, leading to 
di-jet or top-pair final states. This allows for some of the key interactions involved in our baryogenesis mechanism to be tested in high energy experiments.  In particular, we will show that a future hadron collider at a center of mass energy $\sqrt{s}=100$~TeV will be able to discover the new scalars up to $\ord{10}$~TeV masses.  As will be discussed, flavor data provide important constraints on this model and can play a complementary role in probing the parameter space relevant to baryogenesis.

We will next provide a summary of the main features of baryogenesis in the Higgs Troika model.

\section{Baryogenesis from a Higgs Troika} 

\begin{figure}[t]
\centering
\includegraphics[width=\columnwidth]{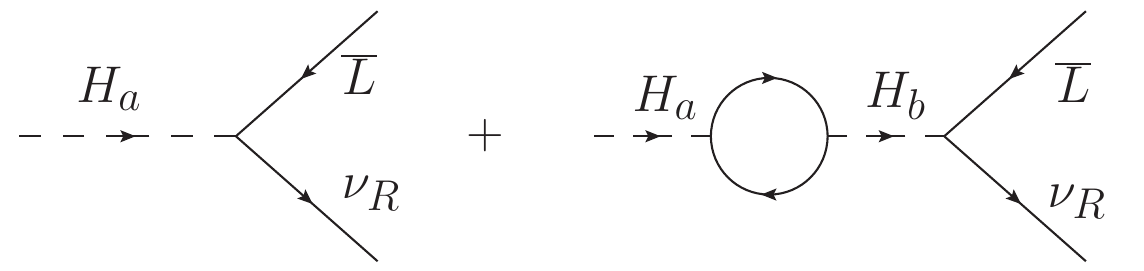}
\caption{Tree and one-loop diagrams for lepton number asymmetry from heavy Higgs decays with $a,b=2,3$.  Other one-loop processes that are not enhanced by scalar mass degeneracy are omitted.}
\label{fig:Feyn-Diag}
\end{figure}

Here, we will outline some of the key features of the Higgs Troika baryogenesis mechanism and refer the interested reader for more details to Ref.~\cite{Davoudiasl:2019lcg}.  We will also clarify where our present work deviates from the assumptions adopted in that reference, though the main ideas are largely the same.  In the Troika framework, the observed BAU is generated through the decays of heavy Higgs fields, with masses $\gsim 1$~TeV, from interference of tree and loop processes, some of which have been illustrated in Fig.\ref{fig:Feyn-Diag}.  In what follows, the three Higgs fields are denoted by 
$H_a$, where $a=1,2,3$, and their interactions with fermions are
\beq
\sum_{a=1}^3\lambda^a_u \tilde H_a^* \bar Q \,u + \lambda^a_d H_a^* \bar Q\, d + 
\lambda^a_\nu \tilde H_a^* \bar L \,\nu_R + \lambda^a_\ell H_a^* \bar L \,\ell\,,
\label{Yukawa}
\eeq   
where fermion generation indices are suppressed.  We will assume that $H_1$ is the SM-like doublet giving rise to all fermion and gauge boson masses, and contains the 
125~GeV boson discovered at the LHC.  Here $H_a$, $a=2,3$, are assumed not to play any role in fermion and gauge boson mass generation, which is a different assumption from that adopted in Ref.~\cite{Davoudiasl:2019lcg}, as mentioned earlier.

Note that we need at least two Higgs doublets  
to get a non-zero CP violating phase, necessary for generating an asymmetry $\eps_a$ in the decay of the doublet $H_a$.  However, given our assumption that $H_1$ gives mass to {\it Dirac} neutrinos, its coupling to the $\bar L\nu_R$ final state, with $\nu_R$ a right-handed neutrino, is very small $\sim 10^{-12}$.  This 
obviously makes it impossible to generate a BAU at the level in \eq{obs-BAU} if $H_1$ is required to contribute.  Therefore, we see that in this case, having two additional fields, $H_{2,3}$ is a requirement for successful baryogenesis, making a Higgs Troika a well-motivated setup.  We will take the heavy Higgs fields to be roughly degenerate in mass, at $\sim$ 5--10\% level, in which case the 1-loop ``bubble'' diagram shown in 
Fig.\ref{fig:Feyn-Diag} dominates over other 1-loop contributions.  

The light particles running in the loop 
could be any of the fermions that couple to $H_a$, however since we will assume that the heavy doublets have $\gsim 0.1$ couplings to SM quarks, they will end up giving the dominant 1-loop contributions to $\eps_a$; other fermion couplings are taken to be smaller by assumption and to satisfy constraints that we will discuss below.  In Refs.~\cite{Egana-Ugrinovic:2018znw,Egana-Ugrinovic:2019dqu}, a flavor framework, dubbed ``Spontaneous Flavor Violation (SFV),'' was proposed to avoid tree level flavor changing neutral currents (FCNCs).  The SFV framework allows for significant couplings of new Higgs states to light quarks in two-Higgs doublet models (2HDMs).  We will adopt this framework, extended to apply to our 
Troika model, as a general guide for the couplings of $H_{2,3}$ to the quark sector of the SM.  We will focus on the ``up-type'' SFV model \cite{Egana-Ugrinovic:2018znw,Egana-Ugrinovic:2019dqu}.    

Let us consider some of the general constraints that apply to the couplings of $H_a$.  An important consideration is to make sure that the asymmetry generated through the decay of the Higgs fields is not 
washed out by scattering processes that allow $f \, \bar f \leftrightarrow \bar \nu_R \,L$, with $f$ any fermion in the SM.  Let us denote the coupling of $H_a$ to $f$ by $\lambda_f^a$.  Following the arguments in Ref.~\cite{Davoudiasl:2019lcg}, inefficient washout by $H_1$ at temperatures 
$T_* \sim 100$~GeV - characteristic of electroweak symmetry breaking (EWSB) - demands 
\beq
\lambda_\nu^1 \,\lambda_f^1 \lsim 10^{-8}\,.
\label{H1-washout}
\eeq  
With our assumption of Dirac neutrino masses from $v\equiv \sqrt{2} \vev{H_1}=246$~GeV, we have 
$\lambda_\nu^1\sim 10^{-12}$ and the above relation is trivially satisfied.  

The decays of $H_{2,3}$ are assumed to be {\it out-of-equilibrium}, that is at temperatures below their mass, hence satisfying Sakharov condition (iii) above.  The washout condition for 
the heavy states then has dependence on their masses $m_a$ which can suppress thermal processes 
$\sim m_a^{-4}$ \cite{Davoudiasl:2019lcg}.  For the rest of this paper, $a=2,3$ unless explicitly specified otherwise.  For a reheat temperature near $T_*=100$~GeV, we find  
\beq
\lambda_\nu^a \,\lambda_f^a  \lsim 2.1 \times 10^{-4} \left(\frac{m_a}{10~\text{TeV}}\right)^2\,,
\label{Ha-washout}
\eeq 
assuming that one flavor of quarks and leptons dominates the underlying processes, through their couplings to $H_a$. In the Appendix, we will outline the derivation of this bound, which is quite consistent with the order-of-magnitude expectation \cite{Davoudiasl:2019lcg}. 

We now consider requirements on couplings from generation of a sufficiently large asymmetry $\eps_a$.  
As will be the case later, let us assume that the largest couplings of $H_a$ are to quarks $q\in \{Q,u,d\}$, as given in \eq{Yukawa}.  Quarks will then be the fermions dominating the loop in Fig.~\ref{fig:Feyn-Diag} and setting the width $\Gamma_a$ of the heavy scalars through $H_a\to \bar q \,q$, given by 
\beq
\Gamma_a = \sum_q\frac{3|\lambda_q^a|^2}{16 \pi}  m_a\,.
\label{Gamma}
\eeq  
To find the regions of parameter space compatible with successful baryogenesis, we follow the formalism developed in Ref.~\cite{Davoudiasl:2019lcg} and now include the width of $H_b$~\cite{Pilaftsis:1997jf}.  In the on-shell renormalization scheme, the dispersive part of the bubble diagram in Fig.~\ref{fig:Feyn-Diag} does not contribute to the decay~\cite{Pilaftsis:1997jf}.     Hence, we only include the absorptive piece of the bubble diagram and the asymmetry parameter that governs baryogenesis generated by the diagrams in Fig.~\ref{fig:Feyn-Diag} is
\begin{eqnarray}
\varepsilon_a = \frac{1}{8\pi}\frac{(m_b^2-m_a^2)m_a^2}{(m_b^2-m_a^2)^2 + m_b^2\Gamma_b^2}\frac{\sum_{f=q} N_{c,f}\text{Im}\left({\rm Tr}^{ba}_\nu {\rm Tr}^{ba*}_{f}\right)}{\sum_{f=q}N_{c,f}{\rm Tr}^{aa}_f}\label{eq:eps}
\end{eqnarray}
where ${\rm Tr}^{ba}_f={\rm Tr}[\lambda^{b\dagger}_{f}\lambda^a_f]$.

A rough schematic estimate for $\eps_a$ is then given by 
\beq
\eps_a \sim \frac{(\lambda_\nu^a)^2}{8 \pi} \sin\theta_f\,,
\label{eps-est}
\eeq
where we have assumed that $H_2$ and $H_3$ have the same couplings to fermions.  In the above, $\sin\theta_f \sim 1$ is assumed and represents the physical CP violating phase contained in the rephasing invariant quantity $\text{Im}({\rm Tr}^{ba}_\nu {\rm Tr}^{ba*}_{f})$ in \eq{eq:eps}, 
necessary to achive an asymmetry.  A short analysis 
\cite{Davoudiasl:2019lcg}, which we recap in the Appendix, can show that a non-thermal production mechanism for $H_a$, based on decay of a heavy modulus of mass $m_\Phi\gsim 2 m_a$, implies that we need 
\beq
\eps_a \gsim 3.4 \times 10^{-8} \left(\frac{m_\Phi}{\text{20~TeV}}\right)\,.
\label{eps-mPhi}
\eeq
Then, \eq{eps-est} requires $\lambda_\nu^a\gsim 10^{-3}$.  Hence, the relation (\ref{Ha-washout}) can be roughly satisfied for $\lambda_\nu^a\sim 10^{-3}$ and $\lambda_q^a\sim 0.1$, with $m_a \sim 10$~TeV, and we could have significant couplings of heavy Higgs states to quarks in the context of the Troika baryogenesis.  

Note that the constraint from \eq{eps-est} can be further relaxed for a modest degeneracy $m_2 \approx m_3$ \cite{Davoudiasl:2019lcg}, which we will assume in our phenomenological study.  Hence, the above conclusions on the size of the quark couplings to $H_a$ can be deemed fairly conservative.  As will be illustrated later, the above interactions allow for a test of the baryogenesis mechanism, through discovery of the Higgs states, up to masses $\gsim 10$~TeV at a future 100 TeV $pp$ collider.    
\section{Flavor Model}

As mentioned earlier, we extend the up-type SFV 2HDM developed in Refs.~\cite{Egana-Ugrinovic:2018znw,Egana-Ugrinovic:2019dqu} for our Yukawa scheme. In the up-type model, the second Higgs doublet couples to up-type quarks proportionally to their SM Yukawa couplings, with a proportionality constant $\xi$. The coupling to down-type quarks, on the other hand, is arbitrary but diagonal in the mass basis, with $\kappa_d$, $\kappa_s$, and $\kappa_b$ denoting the couplings to down, strange, and bottom quarks, respectively. The second doublet also couples to charged leptons with a strength proportional to their SM Yukawa couplings, with a proportionality constant $\xi^\ell$.

We adopt this up-type SFV 2HDM flavor scheme for both the second and third Higgs doublets. However, we also wish to accommodate the Higgs doublets coupling to neutrinos. For this purpose, we additionally add in arbitrary mass-diagonal couplings to neutrinos in a similar fashion to how the Higgs doublets couple to down-type quarks. To summarize, the Yukawa matrices for the heavy Higgs doublets $H_{2,3}$ in our model would be given most generally by

\begin{equation}
\label{eq:flavorscheme}
\begin{aligned}
\lambda^{2,3}_u &= \xi \lambda^1_u 
\\
\lambda^{2,3}_d &= \mathrm{diag}( \kappa_{d},\, \kappa_{s},\, \kappa_{b} ) 
\\
\lambda^{2,3}_\ell &= \xi^\ell \lambda^1_\ell 
\\
\left| \lambda^{2,3}_\nu \right| &= \mathrm{diag}( \kappa_{\nu_1},\, \kappa_{\nu_2},\, \kappa_{\nu_3} ) 
\end{aligned}
\end{equation}
where $\lambda^1_u$ ($\lambda^1_\ell$) is the up-type quark (charged lepton) Yukawa matrix for the doublet $H_1$. SFV is a specific realization of general flavor alignment~\cite{Gatto:1978dy,Gatto:1979mr,Sartori:1979gt,Penuelas:2017ikk,Botella:2018gzy,Rodejohann:2019izm}.  So that our modification to the up-type SFV 2HDM will be minimally invasive, most of the couplings in Eq.~(\ref{eq:flavorscheme}) will be the same or similar for $H_2$ and $H_3$. We will put all the new sources of CP violation into the neutrino Yukawa couplings of $H_2$ and $H_3$. These neutrino couplings $\lambda^{2,3}_\nu$ from Eq.~(\ref{eq:flavorscheme}) will have arbitrary CP violating phases; generating an asymmetry requires a mismatch in the phases between $\lambda^{2}_\nu$ and $\lambda^{3}_\nu$, as can be seen from Eq.~(\ref{eq:eps}). 

The CP violating phases can in principle contribute to the electron electric dipole moment (EDM), $d_e$.
 We note that the analysis of the electron EDM~\cite{Pilaftsis:1997dr} for neutral scalar diagrams shows that $d_e$ is proportional to the mass difference between the CP even and odd scalars in the Higgs doublets.  In our scenario we assume that this mass difference is zero; hence, the electron EDM that is generated is vanishingly small. The one-loop charged Higgs diagrams will involve a neutrino mass insertion~\cite{Bowser-Chao:1997kjp}, so the resulting EDM will be highly suppressed; one can show that the two-loop contributions in our scenario are quite small~\cite{Barr:1990vd,Bowser-Chao:1997kjp}. Any electron EDM in our model will be several orders of magnitude below the current electron EDM 
bound of $d_e < 1.1 \times 10^{-28}$~$e$ cm \cite{Andreev:2018ayy}. Hence, the electron EDM is not constraining.

\begin{figure*}[t]
\subfigure[]{ \includegraphics[width=\columnwidth]{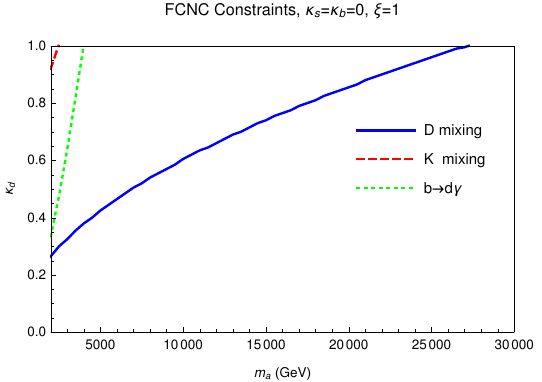}}
\subfigure[]{ \includegraphics[width=\columnwidth]{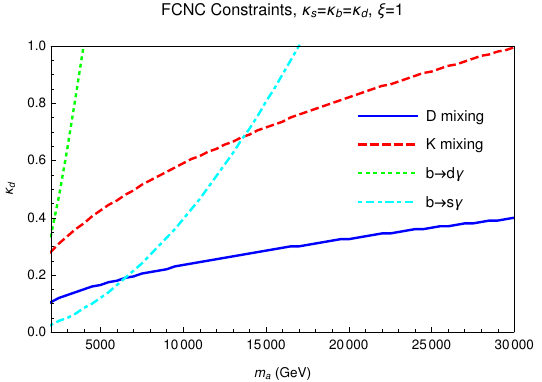}}
\subfigure[]{ \includegraphics[width=\columnwidth]{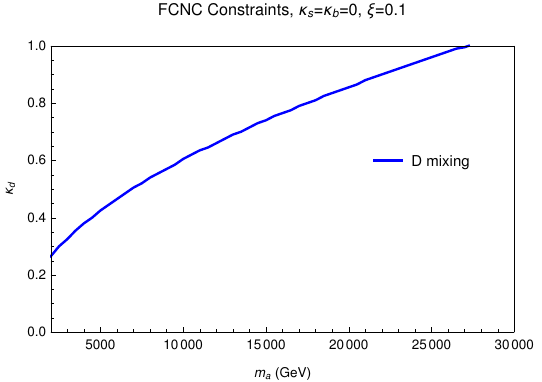}}
\subfigure[]{ \includegraphics[width=\columnwidth]{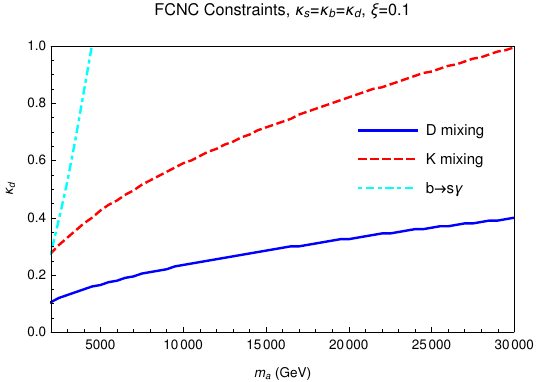}}
\caption{Upper bounds from neutral meson mixing and flavor-changing decays on the coupling of $H_{2,3}$ to the down quark, $\kappa_d$, for some sample parameter assignments of (a) $\kappa_s=\kappa_b=0$, $\xi=1$; (b) $\kappa_s=\kappa_b=\kappa_d$, $\xi=1$; (c) $\kappa_s=\kappa_b=0$, $\xi=0.1$; (d) $\kappa_s=\kappa_b=\kappa_d$, $\xi=0.1$. The curves represent bounds from $D$ meson mixing (blue solid), $K$ meson mixing (red dashed), $b \to d\gamma$ decays (green dotted), and $b \to d\gamma$ decays (cyan dot-dashed).}
\label{fig:flavorbounds}
\end{figure*}

The SFV scheme suppresses FCNC processes, but experimental bounds are still constraining, especially for lighter Higgses. The relevant experimental bounds come from the flavor-changing decays $b \to d\gamma$ \cite{Crivellin:2011ba} and $b\to s\gamma (\ell^+\ell^-)$ \cite{Capdevila:2017bsm}, as well as neutral meson mixing for $K-\bar{K}$\cite{Bona:2007vi}, $B_d-\bar{B}_d$ \cite{Bona:2016bvr}, $B_s-\bar{B}_s$ \cite{Bona:2016bvr}, and $D-\bar{D}$ \cite{Aaij:2019jot}. We use the formulas presented in Ref.~\cite{Egana-Ugrinovic:2019dqu} to calculate the contributions to these different processes in our model and have checked that we reproduce their results. Assuming $H_2$ and $H_3$ have identical masses and Yukawa couplings, and taking the alignment limit, we show the relevant limits that these experimental bounds place on the couplings in Fig.~\ref{fig:flavorbounds}.

For the multi-TeV masses and Yukawa couplings that we consider, the most constraining measurements are $D-\bar{D}$ mixing and sometimes $b\to s\gamma (\ell^+\ell^-)$ for lighter masses. The $D$ meson mixing constraints are largely independent of $\xi\lesssim 1$ due to the dominant contribution of $\kappa_d,\kappa_s,\kappa_b$ via charged Higgs loops. The flavor-changing decay $b \to d\gamma$ is only the strongest constraint below the mass region we are interested in, but is still close to being competitive for masses of around 2 TeV. The constraints for $B_s-\bar{B}_s$ and $B_d-\bar{B}_d$ mixing are constraining only for masses lower than what we consider, but $K-\bar{K}$ mixing constraints are within about a factor of 2 of the stronger $D-\bar{D}$ mixing and $b\to s\gamma (\ell^+\ell^-)$ bounds when multiple down-type Yukawa couplings are non-zero. Improved measurements for some of these processes, in particular $D-\bar{D}$ mixing at LHCb, and also perhaps $K-\bar{K}$ mixing and the $b \to s\gamma$ and $b \to d\gamma$ flavor-changing decays, may provide discovery channels for our model in the future. There is also much work going into calculating the SM predictions to greater precision; see Ref.~\cite{Lehner:2019wvv} for a recent discussion of the flavor physics prospects of future lattice QCD improvements in combination with current and future experiments.

\section{Collider Signatures}
\begin{figure*}
\subfigure[]{  \includegraphics[width=0.45\textwidth]{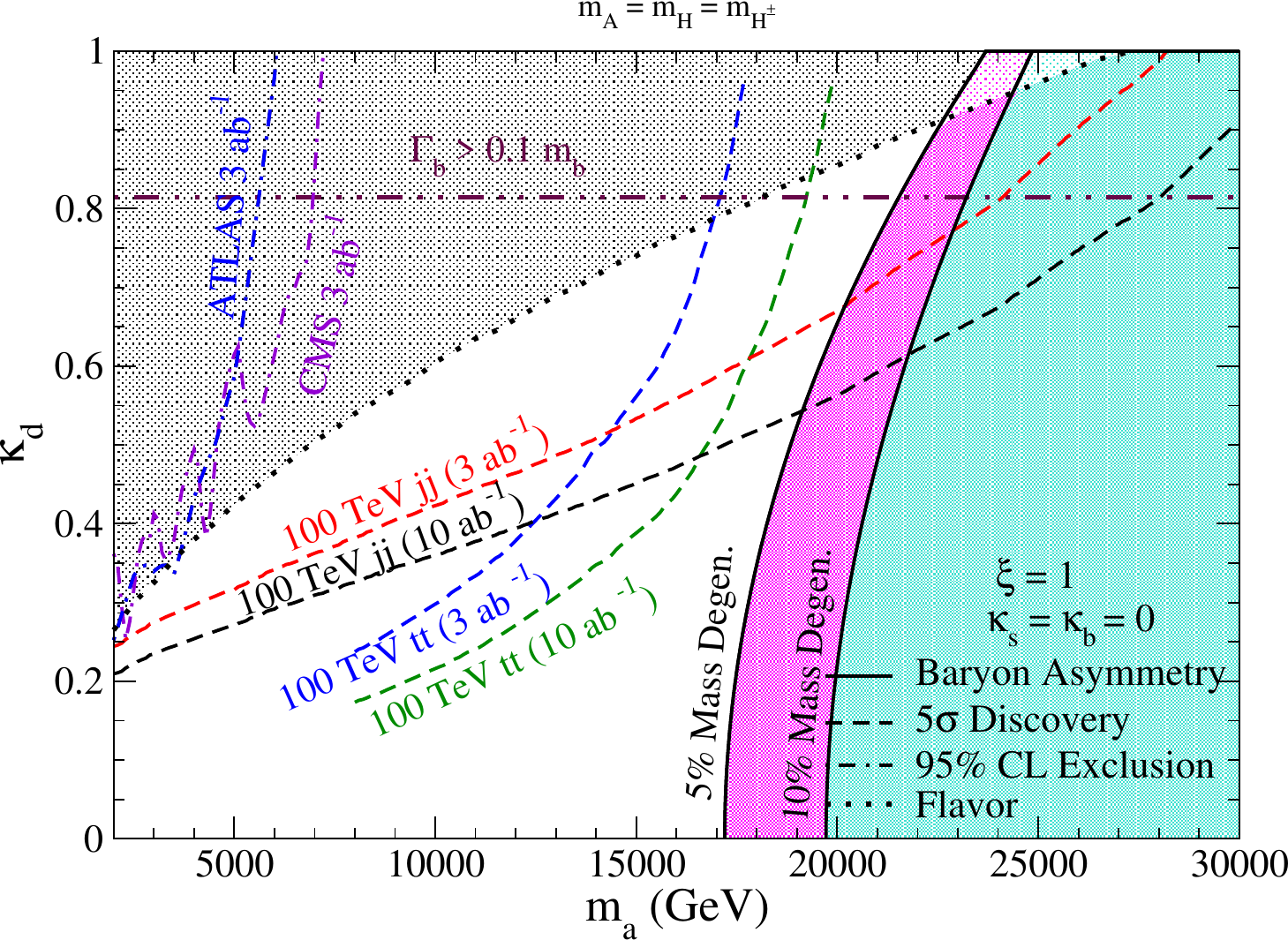}}
\subfigure[]{  \includegraphics[width=0.45\textwidth]{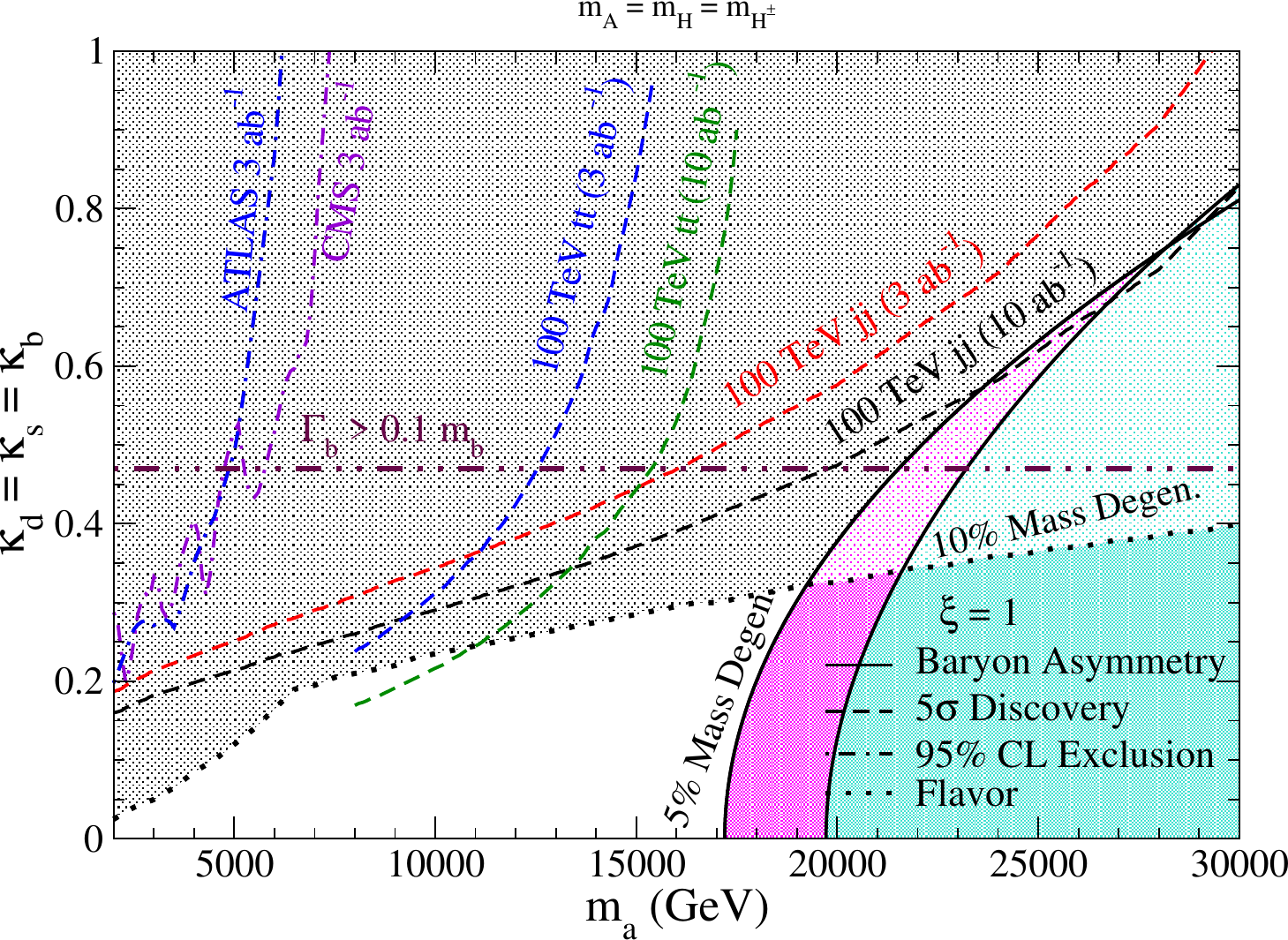}}
\subfigure[]{  \includegraphics[width=0.45\textwidth]{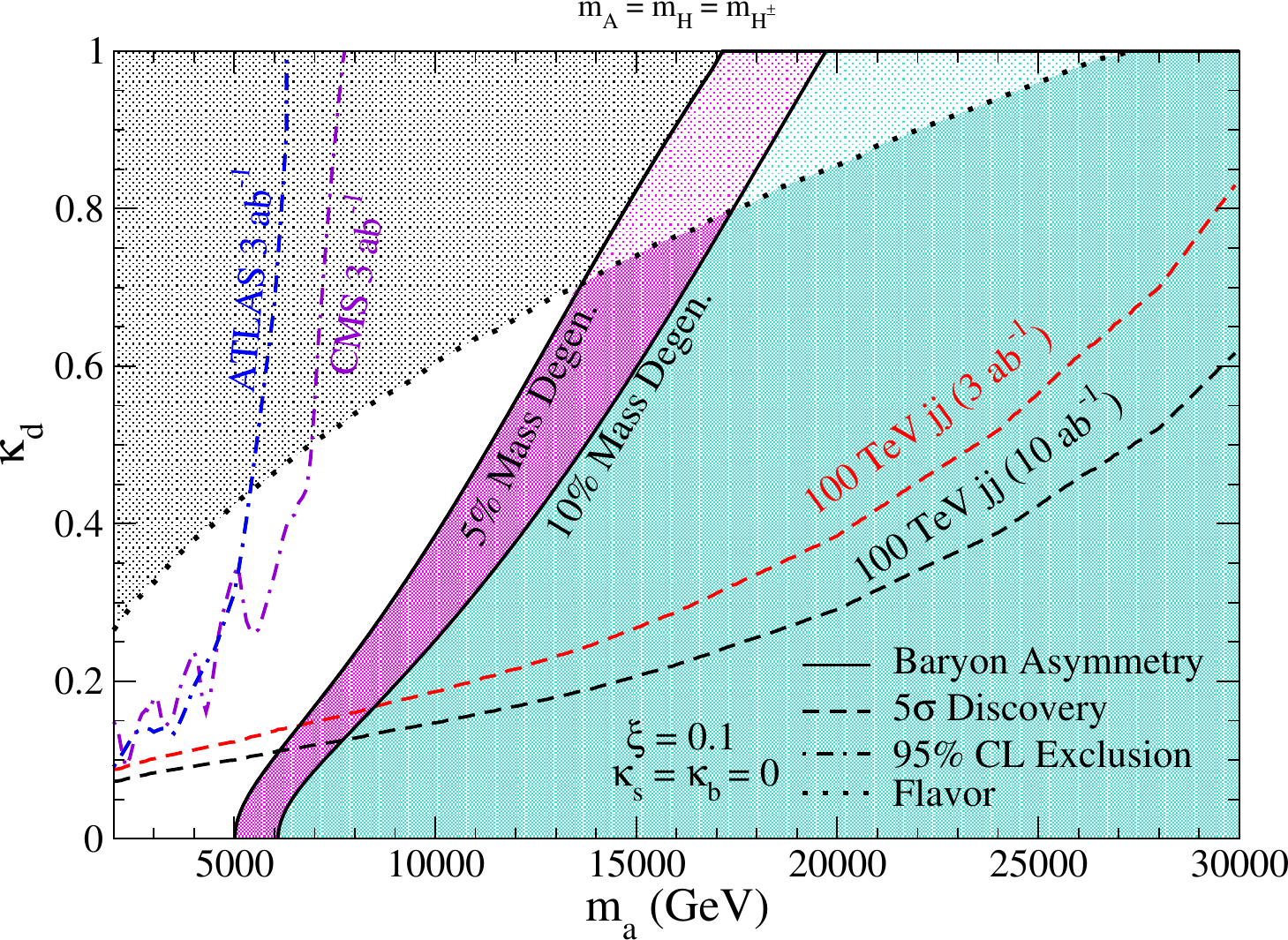}}
\subfigure[]{  \includegraphics[width=0.45\textwidth]{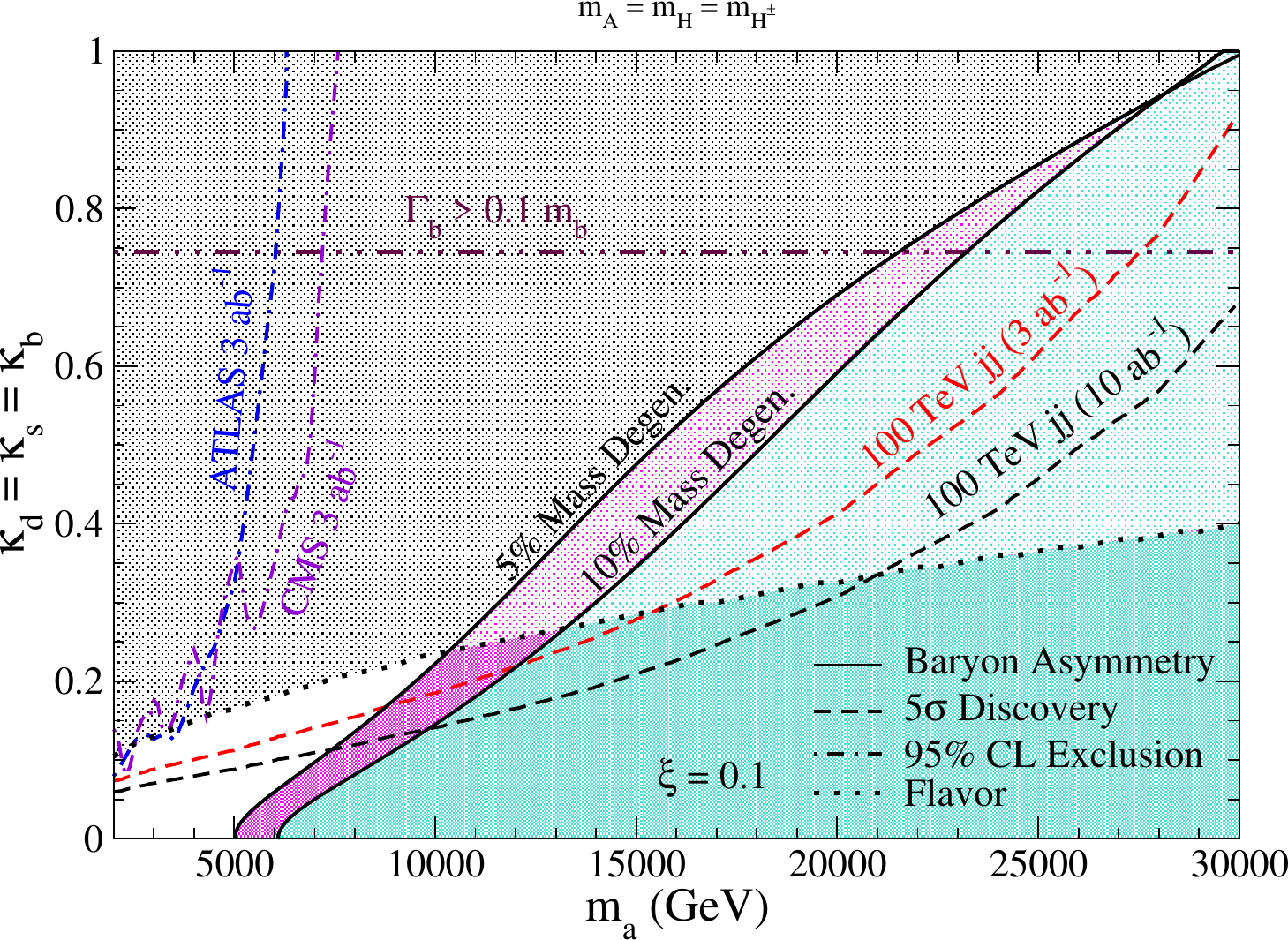}}
\subfigure[]{  \includegraphics[width=0.45\textwidth]{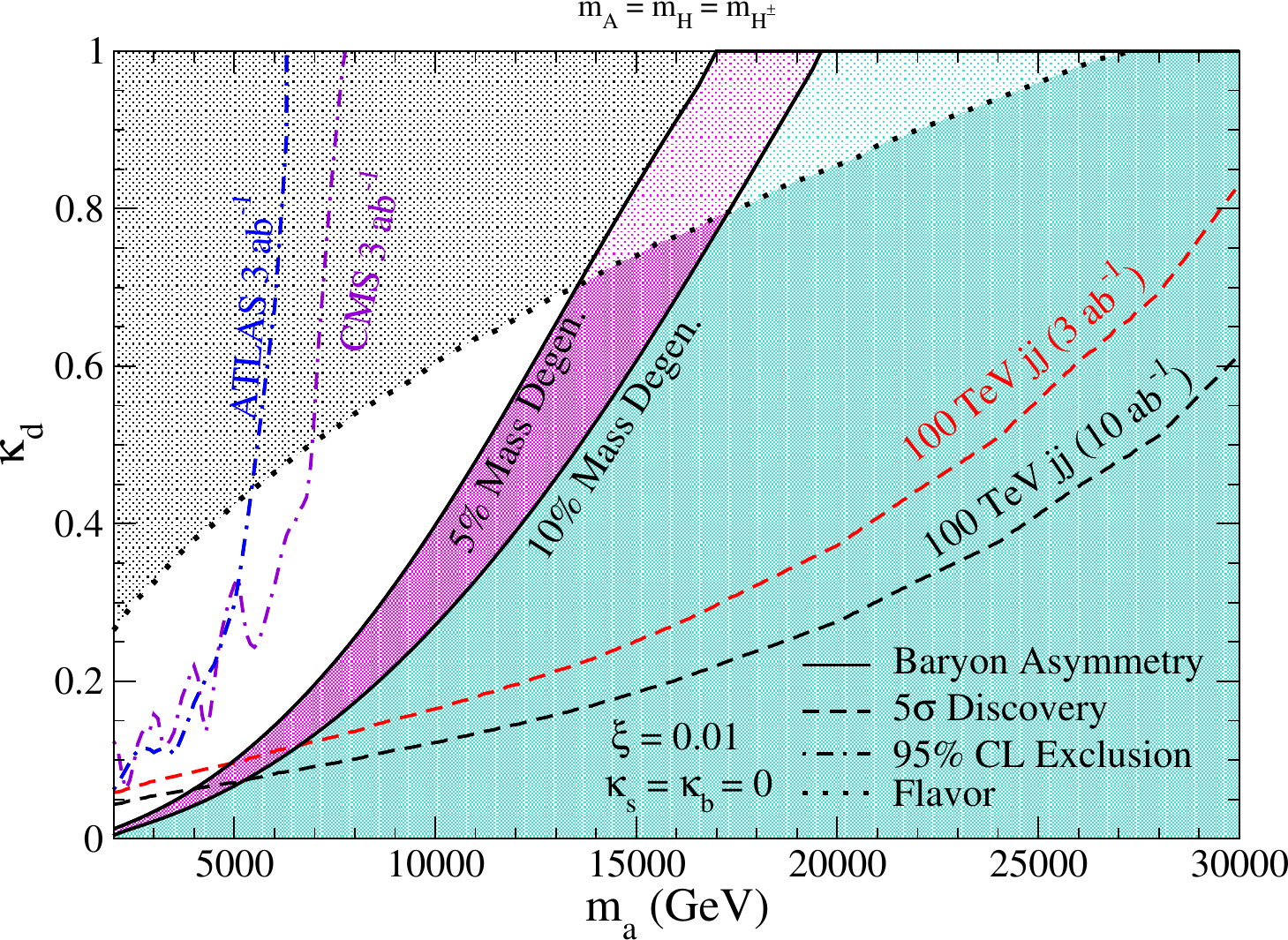}}
\subfigure[]{  \includegraphics[width=0.45\textwidth]{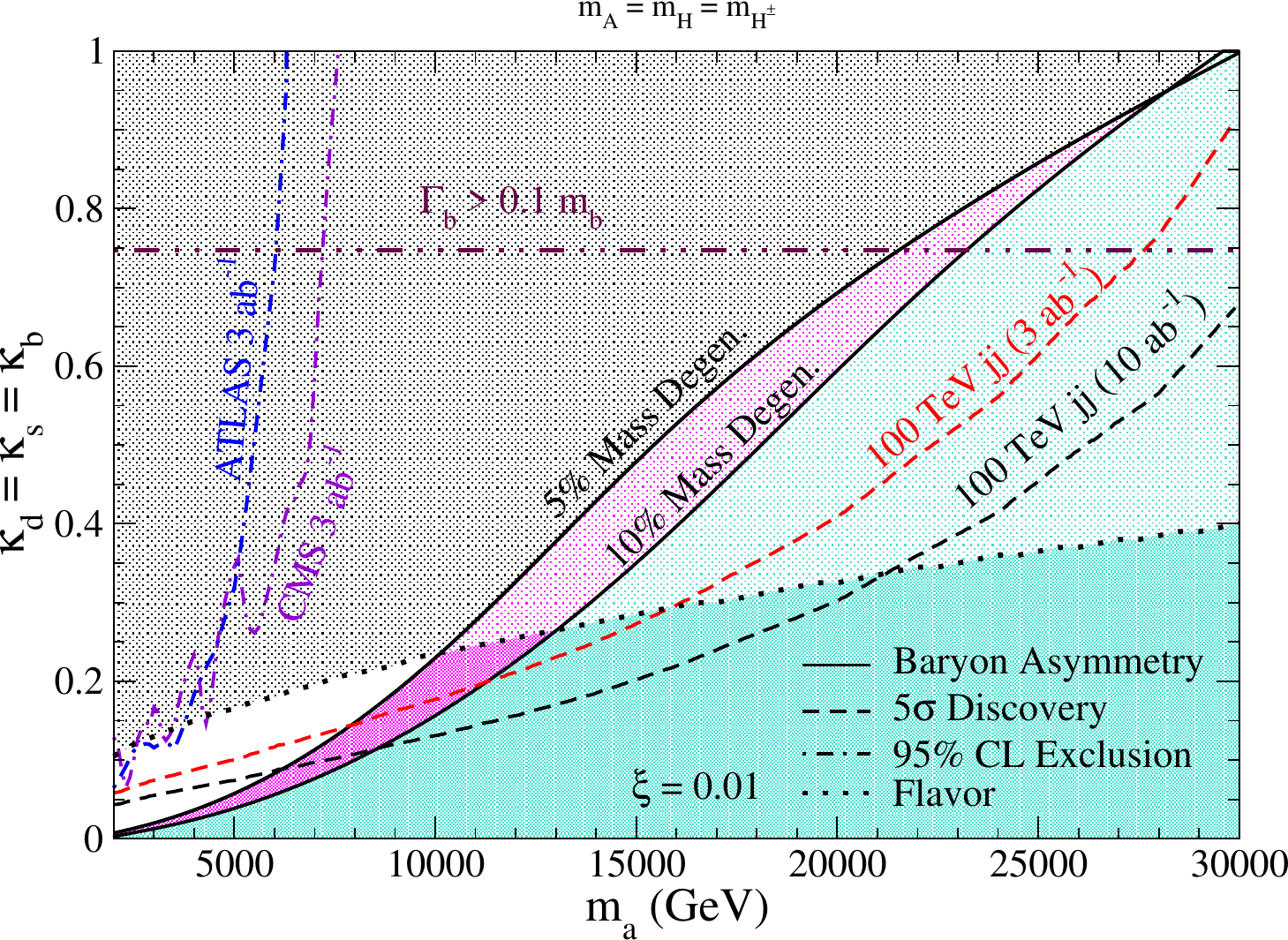}}
  \caption{Regions above dashed lines are discoverable at 5$\sigma$ in the dijet channel at a 100 TeV collider with 3 ab$^{-1}$ (red) and 10 ab$^{-1}$ (black), and the $t\bar{t}$ channel with 3 ab$^{-1}$ (blue) and 10 ab$^{-1}$ (green).  Regions above dot-dashed lines can be excluded at 95\% at ATLAS (blue) and CMS (violet) with 3 ab$^{-1}$.  The regions below the solid lines and color shaded are compatible with successful baryogenesis with a 5\% mass degeneracy $m_a/m_b=0.95$ (magenta) and a 10\% mass degeneracy $m_a/m_b=0.9$ (turquoise).  Regions above the dotted lines and either grey or light color shaded are excluded by the flavor bounds presented in Fig.~\ref{fig:flavorbounds}.  Regions above the maroon dot-dot-dashed lines have width to mass ratios $\Gamma_b/m_b$ larger than 10\%; we expect our approximation of the decay asymmetry to be good below this line. These are shown in the $\kappa_d$ vs. Higgs doublet mass, $m_a$, plane for (a,c,e) $\kappa_s=\kappa_b=0$; (b,d,f) $\kappa_d=\kappa_s=\kappa_b$; (a,b) $\xi=1$; (c,d) $\xi=0.1$; and (e,f) $\xi=0.01$.\label{fig:exc}}
\end{figure*}

As mentioned previously, with the couplings in Eq.~(\ref{eq:flavorscheme}) it is possible to resonantly produce the additional heavy Higgs bosons at hadron colliders via direct couplings to quarks~\cite{Egana-Ugrinovic:2018znw,Egana-Ugrinovic:2019dqu}.  This is unlike traditional Higgs processes where the Higgs couples so weakly to the light quarks it must either be produced via loop level gluon and quark processes, or in multi-body final states.  Resonant production with light quark initial states greatly enhances the hadron collider reaches.

Assuming the additional heavy Higgses do not mix with the 125 GeV Higgs (``alignment limit''), they will not decay to gauge bosons or the SM-like Higgs pairs.  Hence, once a heavy Higgs boson is produced it will dominantly decay into jets or top quarks.  We project bounds from the LHC with 3 ab$^{-1}$ and discovery potential at a 100 TeV pp collider with 3 ab$^{-1}$ and 10 ab$^{-1}$ in the dijet channel.  To generate production cross sections the couplings of Eq.~(\ref{eq:flavorscheme}) are implemented in \texttt{MadGraph5\_aMC@NLO}~\cite{Alwall:2014hca} via \texttt{FeynRules}~\cite{Christensen:2008py,Alloul:2013bka}.  The results of these extrapolations are shown in Fig.~\ref{fig:exc}.  The results at the HL-LHC are accomplished via a simple root luminosity scaling of current dijet bounds from ATLAS~\cite{Aad:2019hjw} and CMS~\cite{Sirunyan:2019vgj}.  Both ATLAS and CMS present their results as limits on cross section times branching ratio times acceptance.  We calculated acceptances using parton level acceptance cuts in~\texttt{MadGraph5\_aMC@NLO}.  

For the 100 TeV collider 5$\sigma$-discovery reach, we re-interpreted the $Z_B$ bounds projected in Ref.~\cite{Golling:2016gvc}.  Using the same methods above, we implemented the $Z_B$ model in~\texttt{MadGraph5\_aMC@NLO}, and translated the gauge coupling vs. mass limits into dijet cross sections limits.  The limits and discovery reaches presented in Fig.~\ref{fig:exc} assume one Higgs doublet with equal charged Higgs mass, $m_{H^+}$; scalar mass, $m_H$; and pseudoscalar mass $m_{A}$.  With the equal mass and alignment assumptions, the heavy Higgs doublets give vanishing contributions to the $S,~T$, and $U$ oblique parameters~\cite{Peskin:1990zt,Peskin:1991sw,Barbieri:2006dq,Haber:2010bw,Ahriche:2015mea}. We would like to make two observations about the equal mass assumption:
\begin{enumerate}
\item The mass differences between the heavy scalar, pseudoscalar, and charged Higgses originate from EWSB.  Hence, the mass differences are proportional to the EW vev.  For a 10 TeV Higgs at order one couplings in the scalar potential, this is roughly a $0.02-0.03\%$ mass difference.  The decay of a 10 TeV particle will result in jets with energy of $\sim 5$~TeV.  The jet energy resolution of a jet with energy $5$ TeV at a future circular hadron collider is $\sim 3\%$~\cite{Benedikt:2018csr}.  Hence, the reconstructed di-jet resonance is expected to have a mass resolution of $\sim 6\%$, well above the theoretically expected mass difference of $0.02-0.03\%$.  Hence, the resonances would be an overlapping distribution, and our assumption of equal masses is not far from what may realistically be expected at a collider.
\item The subject of interference between degenerate scalars and pseudoscalars is an interesting one~\cite{Pilaftsis:1997dr}.  In the case they are CP eigenstates, they do not interfere and their rates add incoherently.  However, in the model we study, there is CP violation.  Hence, the scalar and pseudoscalar can mix via fermion loops.  If this mixing is large, interference effects must be accounted for when calculating the production cross section at colliders.  These interference effects can be large and negative~\cite{Pilaftsis:1997dr}, decreasing the rate below the na\"{i}ve factor of two.  However, in our case we can place the CP-violating phase in the neutrino Yukawa couplings.  Since the neutrinos couple very weakly to all Higgses, the loop induced scalar-pseudoscalar coupling is expected to be negligible.  Hence, in our calculation of rates we take the optimistic scenario and assume the scalar and pseudoscalar add incoherently.  If there is large mixing between the scalar and pseudoscalar, a more careful calculation is warranted.
\end{enumerate}
Finally, in calculating rates at colliders we use the couplings to physical Higgs bosons in the alignment limit as given in Table 7 of Ref.~\cite{Egana-Ugrinovic:2019dqu}\footnote{We account for a typo in v2 of Ref.~\cite{Egana-Ugrinovic:2019dqu} and include a missing $1/\sqrt{2}$ normalization for the neutral scalars in Table 7. }.

Assuming all quarks contribute to the washout condition, using the flavor structure under consideration given in \eq{eq:flavorscheme}, and applying the washout condition in Eq.~(\ref{Ha-washout}), we find
\begin{eqnarray}
\varepsilon_a &\lesssim& 1.8\times10^{-9}\left(\frac{m_a}{10 {\rm TeV}}\right)^4\frac{(m_b^2/m_a^2-1)}{(m_b^2/m_a^2-1)^2 + m_b^2\Gamma_b^2/m_a^4}\nonumber\\
&&\times \frac{1}{\kappa_d^2+\kappa_s^2+\kappa_d^2+\xi^2}\,.\label{eq:epsnum}
\end{eqnarray}
In the above expression, we have implicitly assumed that there is a physical phase $\theta_f\neq 0$, necessary to generate an asymmetry and originating from $\text{Im}({\rm Tr}^{ba}_\nu {\rm Tr}^{ba*}_{f})$ in \eq{eq:eps}, with $|\sin \theta_f| \lsim 1$.  We note that as $m_a\to m_b$, the width of the intermediate $H_b$ in Fig.~\ref{fig:Feyn-Diag} will regulate the growth of $\eps_a$ and further mass degeneracy will not enhance the asymmetry and dominance of the 1-loop process shown in the figure.     

Since we consider Higgs masses up to 30 TeV, the mass of the modulus that decays into the Higgs bosons to create the baryon asymmetry must satisfy $m_\Phi\gsim 60$ TeV.  From Eq.~(\ref{eps-mPhi}), this translates to a requirement that $\varepsilon_a\gtrsim 10^{-7}$; the shaded regions in Fig.~\ref{fig:exc} are derived using this baryogenesis requirement.  For a given Higgs mass and $\xi$, the washout condition in Eq.~(\ref{Ha-washout}) sets a maximum neutrino coupling, while $\varepsilon_a \gtrsim 10^{-7}$ sets a minimum. As $m_a$ increases, the coupling $\lambda^a_\nu$ compatible with washout increases and successful baryogenesis can occur.  Hence, for fixed $\xi$ there is a minimum $m_a$, typically in the multi-TeV region, required to simultaneously satisfy both wash-out and asymmetry parameter requirements.  For smaller $\xi$ the washout condition is relaxed and smaller Higgs masses are allowed.

Despite needing multi-TeV Higgses, as can be clearly seen in Fig.~\ref{fig:exc}, a 100 TeV pp collider has great potential to discover this scenario.  As $\xi$ becomes smaller, the branching ratios into top quark final states are suppressed, enhancing the dijet cross section.  Hence, the discovery potential is greater for smaller $\xi$.  Reference~\cite{Golling:2016gvc} also has projected bounds on $t \bar t$ resonances at a 100 TeV.  For higher values of $\xi\sim 1$, the discovery reach in the $t \bar t$ channel is greater than the dijet channel for Higgs masses below $\sim 11-17$~TeV. However, as seen in the Figs.~\ref{fig:exc}(a,b), the dijet searches are still more sensitive to the regions of parameter space that can accommodate baryogenesis.  The top pair bounds are not relevant for $\xi=0.01$ and $0.1$, and hence are not shown.

We also superimpose the most stringent flavor bounds from Fig.~\ref{fig:flavorbounds} onto the collider and baryogenesis constraints in Fig.~\ref{fig:exc}.  For $\kappa_s=\kappa_b=0$, a 100 TeV collider always has a greater reach than the searches in flavor physics.  However, this is not true for $\kappa_d=\kappa_s=\kappa_b$.  In this case, for $\xi=1$ the flavor constraints rule out the parameter regions a 100 TeV collider could discover (with our assumptions, as further discussed below).  For $\xi=0.1$ and $\xi=0.01$, there are allowed baryogenesis compatible regions that a 100 TeV collider could discover for $m_a\lesssim 15-20$~TeV.  From this discussion, it is clear that measurements of flavor observables and searches at colliders are sensitive to complementary regions of parameter space.  Improvements in $D-\bar{D}$ constraints are especially sensitive to $\kappa_d=\kappa_s=\kappa_b$, $\xi=1$ and $m_a\gtrsim 15-20$ TeV.  A 100 TeV collider has great potential to discover baryogenesis in this model for most all other parameter regions.

Our calculation of asymmetry parameter depended on the dominance of the bubble diagrams in Fig.~\ref{fig:Feyn-Diag} over possible triangle diagrams.  Hence, we include maroon dot-dot-dashed lines in Fig.~\ref{fig:exc} where the width of $H_b$ is 10\% of its mass.  Below these lines we expect our estimate of $\varepsilon_a$ to be valid.  If a plot does not include this line, then all regions of parameter space are consistent with $\Gamma_b < 0.1\,m_b$.  For the $\kappa_d=\kappa_s=\kappa_b$ cases (the right-hand side of Fig.~\ref{fig:exc}), it appears that our calculation is not valid in many regions of parameter space shown.  However, we would note that the flavor constraints force our model into regions where $\Gamma_b < 0.10\,m_b$, except for a tiny corner of parameter space in Fig.~\ref{fig:exc}(a).  Hence, we expect our results to be robust.

Finally, we note that our dijet projections are conservative.  It has been assumed only one Higgs doublet is searched for.  However, a distinct prediction of our model is that for successful baryogenesis with $m_a\gtrsim \mathcal{O}(1~{\rm TeV})$, there is a second Higgs doublet close in mass.  For larger mass separations, this could generate a signal of multiple distinct dijet resonances.  If the separation between the Higgs bosons masses is less than the detector jet energy resolution but greater than the Higgs widths, our signal may appear as a broad resonance with twice the signal cross section.  This would increase our cross section by up to a factor of two and coupling reach up to a factor of $\sim \sqrt{2}$.  If the Higgs mass separation is less than the Higgs widths, we would expect a coherent enhancement, since the two Higgs doublets have the same quantum numbers.  This would increase our cross section up to a factor of four and coupling reach up to a factor of $\sim 2$.  The precise details of how to search for these scenarios depend  intimately on the Higgs mass spectrum, the magnitude of the quark couplings and Higgs widths, and the resolution of future detectors.    However, even in the conservative scenario considered here, a 100 TeV pp collider has great potential to discover the baryogenesis mechanism we have proposed.

The factor of 2 sensitivity enhancement from constructive coherent interference between the heavy Higgses will increase the sensitivity of ATLAS and CMS to our baryogenesis mechanism.  To satisfy Eq.~(\ref{eq:epsnum}), $\varepsilon_a\gtrsim10^{-7}$, and to have the mass difference between the heavy Higgs bosons be less than the width, we find the width to mass ratio must be percent level or smaller.  Given that level of degeneracy, ATLAS and CMS may be sensitive to our baryogenesis mechanism for $m_a\lesssim 5-7$~TeV and $\xi\lesssim 0.1$.
\section{Conclusions}

In this work, we considered the Higgs Troika baryogenesis mechanism assuming that the heavy scalar states from new Higgs doublets couple to light quarks with significant strengths.  This setup can avoid large flavor violation effects, assuming the ``Spontaneous Flavor Violation'' framework, which we adapted for our proposal as a general guide for the new Yukawa couplings.  The light quark couplings to the two new Higgs doublets allow for their resonant production.  These interactions are also key components of the proposed baryogenesis mechanism, which favors a hierarchy of masses between the SM-like Higgs and the new doublets to avoid the washout of baryon number.  

We find that direct searches at colliders and indirect searches from flavor physics are sensitive to different regions of parameter space that are consistent with successful baryogenesis.  If only one down-type quark coupling is non-zero, then a 100 TeV collider will be the main discovery channel, and with 3-10 ab$^{-1}$ of integrated luminosity can probe significant regions of the baryogenesis parameter space.  When all down-type quark couplings are non-zero and equal, searches in flavor space are sensitive to our baryogenesis mechanism in regions of parameter space complementary to direct searches at a 100 TeV.   Finally, if the mass difference between the heavy Higgses is less than their widths, the production and decay cross sections are coherently enhanced.  In this case the LHC has the potential to probe baryogenesis in the Higgs Troika model for masses $\lesssim 5-7$~TeV and width to mass ratios at the percent level or smaller.  Hence, current and envisioned future collider experiments as well as improvements in bounds on flavor observables can potentially probe the new states and examine their relevance to the proposed processes for generating the baryon asymmetry of the Universe.

\begin{acknowledgments}
{\bf Acknowledgments:} We thank S. Homiller for helpful communication.  H.D. and M.S. are supported by the United States Department of Energy under Grant Contract DE-SC0012704.  I.M.L. is supported in part by the United States Department of Energy grant number DE-SC0017988.  The data to reproduce the plots are available upon request.
\end{acknowledgments}

\appendix*

\section{Appendix}

\subsection{Washout Rate Estimation}

Here, we outline our approach to the estimation of the lepton number asymmetry $\Delta L$ washout rate.  Since the Troika baryogenesis mechanism relies on the amount of $\Delta L$ generated via heavy Higgs doublet decays, one needs to make sure that its washout is not efficient once the reheat temperature $T_{\rm rh}$ has been established.  For a viable scenario, $T_{\rm rh}$ needs to be at or above the electroweak phase transition temperature $T_* \sim 100$~GeV, in order to have active sphaleron processes necessary to generate the baryon number from $\Delta L$.  

In order to determine the efficiency of the washout rate $\Gamma_{\rm wo}(T)$ at temperature $T$, we will consider the ratio (see, for example, Ref.~\cite{Cui:2011ab})  
\beq
\frac{\Gamma_{\rm wo}(T)}{{\cal H}(T)} = 
\frac{\avsig}{2 \,n_\gamma (T){\cal H}(T)} \prod_i n_i (T)\,,    
\label{wo-rate}
\eeq
where ${\cal H} (T) \approx 1.66\, g_*^{1/2} T^2/\mP$ is the Hubble rate during the radiation dominated era and $g_*$ is the number of relativistic degrees of freedom; the Planck mass is given by $\mP \approx 1.2 \times 10^{19}$~GeV \cite{Tanabashi:2018oca}.  In \eq{wo-rate}, $\avsig$ is the thermally averaged washout cross section, $n_\gamma (T)$ is the photon number density and $n_i(T)$ are the number densities for the initial state particles relevant to the underlying process.  A washout process is deemed ineffective if the condition 
\beq
\frac{\Gamma_{\rm wo}(T)}{{\cal H}(T)}<1
\label{cond}
\eeq
is satisfied.  

To calculate $\avsig$, we will follow the formalism of Ref.~\cite{Srednicki:1988ce}.  We have
\beq
\avsig = \frac{1}{{\bar n_1}{\bar n_2}}
\int \frac{d^3p_1 \,d^3p_2}{E_1\, E_2}\, f(E_1) f(E_2)\, w(s)\,,
\label{avsig}
\eeq  
where initial state quantities are denoted by subscripts \{1, 2\}, assuming $2\to j$ processes; $j\geq 1.$  In the above, $f(E_{1,2})$ denote energy distributions of initial states, and $\bar n \equiv \int d^3 p f(E)$. In the relativistic non-degenerate regime, relevant to our estimate here, we can assume a Boltzmann distribution for the initial states with $f(E) = e^{-E/T}$, to a good approximation \cite{Kolb:1990vq}.  This choice for $f(E)$ yields ${\bar n} = 8 \pi\, T^3$, in the massless limit, which is the case for our calculations below. The function $w(s)$, with $s = (p_1 + p_2)^2$ the center of mass energy squared, is defined by 
\beq
w(s) \equiv \frac{1}{4}\int d\Phi\, \overline{|{\cal M}|^2}\,,
\label{w}
\eeq    
where the Lorentz-invariant phase space is given by  
\beq
d\Phi = (2\pi)^4\, \delta^{(4)} (p_1 + p_2 - \sum_j p_j) \prod_j \frac{d^3p_j}{(2\pi)^3\, 2 E_j}
\label{dPhi}
\eeq
and $\overline{|{\cal M}|^2}$ denotes the squared amplitude for the underlying washout process, averaged and summed over the initial and final state quantum numbers. 

Since the function $w(s)$ is Lorentz-invariant, we can calculate it in any convenient frame and recast the result as a function of $s$.  Then, one may proceed to calculate $\avsig$ in \eq{avsig} using 
\beq
s = M_1^2 + M_2^2 + 2 (E_1 E_2 - |\vec{p_1}| |\vec{p_2}| \cos\zeta)\,, 
\label{s}
\eeq
where $\zeta$ is the angle between initial momenta.  For the washout processes relevant to our work, the initial states are massless: $M_1=M_2=0$. 

In the model presented in this paper, we have assumed that quark couplings to heavy Higgs states $H_a$ are much larger than their couplings to leptons.  Hence, we only consider $\Delta L$-erasing processes that involve at least one quark in the initial state.  Also, we will not consider initial $\nu_R$ states since they are not thermally populated in our scenario, in general.  This is because the SM-like Higgs couplings to neutrinos $\sim 10^{-12}$, for the assumed Dirac neutrino masses, are too weak and we are only interested in parameters for which washout through $H_a$ is ineffective. 

With the above assumptions, the relevant $t$-channel processes we consider are 
$u L \to \nu_R Q$, $\bar Q L \to \nu_R \bar u$, $Q L\to \nu_R d$, and $\bar d L \to \nu_R \bar Q$, where $(Q,L)$ denote (quark, lepton) doublets 
and $(u,d)$ denote up-type and down-type quark singlets, respectively.    
The $s$-channel processes are $\bar d Q \to \nu_R \bar L$ and $\bar Q u\to \nu_R \bar L$.  As we have assumed nearly degenerate $H_a$ states, with the same couplings to various fermions, each of the preceding processes is summed over both contributions.  In our calculation, we consider the case where only one generation of quarks and leptons - with Yukawa couplings $\lambda^a_q$ and $\lambda^a_\nu$, respectively - dominate the washout rate.  For $T_*=100$~GeV and $m_a=10$~TeV, 
we find $\lambda^a_q \lambda^a_\nu \lsim 2.1 \times 10^{-4}$, as presented in \eq{Ha-washout}.

\subsection{Asymmetry from Modulus Decay}

Let us assume a population of $\{H_a, H_a^*\}$ is non-thermally produced through the decays of a heavy modulus $\Phi$ of mass $m_\Phi$; 
$\{H_b,H_b^*\}$ number density is taken to be negligible.  The initial energy 
density of $H_a$ is given by $\rho_a \sim E_a n_a$, where $n_a$ is the $H_a$ number density and we have $E_a \sim m_\Phi/2$.  The decay of $H_a$ contributes to reheating of the Universe and $\rho_a \leq \rho_R$, with $\rho_R$ the radiation energy density; $\rho_R = (\pi^2/30) g_* T^4$, where $g_*$ is the number of relativistic degrees of freedom and $g_* = 106.75$ in the SM.  We define the ratio   
\beq
r \equiv \frac{E_a n_a}{\rho_R}, 
\label{r}
\eeq
which satisfies $r \leq 1$.  The $B-L$ abundance is then given by 
\beq
\frac{n_{B-L}}{s} =  \frac{3\, r \,T_{\rm rh} \, \eps_a}{4\, E_a},
\label{B-L-abund}
\eeq
where $s = (2 \pi^2/45) g_* T^3$ is the entropy density and the reheat temperature is $T_{\rm rh}$.

Since the new Higgs states are assumed heavy compared to $T_{\rm rh}$ (departure from equilibrium), the relevant degrees of freedom are those of the SM and, using the results of Ref.~\cite{Harvey:1990qw}, one has  
\beq
\Delta B = \frac{28}{79}\, \DBL.
\label{BBL}
\eeq 
Given \eq{BBL}, we then have  
\beq
\frac{n_B}{s} = \frac{21}{79} \left(\frac{r \,T_{\rm rh} \, \eps_a}{E_a}\right).
\label{BAU}
\eeq
Using \eq{obs-BAU} and $T_{\rm rh}\sim 100$~GeV, the above then yields \eq{eps-mPhi}.

\bibliography{higgstroika-refs}

\end{document}